\documentstyle[twocolumn,aps,pre,psfig]{revtex}

\newcommand{\so}{SOC }

\newcommand{\ra}{$1/f^\alpha$ noise }

\begin{document}
\draft
\twocolumn[\hsize\textwidth\columnwidth\hsize\csname @twocolumnfalse\endcsname
\title{$1/f^\alpha$ noise from self-organized critical models with uniform driving $^\dagger$}
\author{J\"orn Davidsen \cite{byline} and Heinz Georg Schuster}
\address{Institut f\"ur Theoretische Physik und Astrophysik, Christian-Albrechts-Universit\"at,\\
Olshausenstra\ss e 40, 24118 Kiel, Germany\\
$^\dagger$ accepted by Phys. Rev. E.}
%\date{21 June, 2000}
\maketitle

\begin{abstract}
Using the well-known Olami-Feder-Christensen model as our paradigm, we show how to modify uniform driven self-organized critical models to generate $1/f^\alpha$ noise. Our model can reproduce all the main features of $1/f^\alpha$ noise: (1) $\alpha$ is close to one and does not depend on the dimension of the system. (2) The $1/f^\alpha$ behavior is found for very low frequencies. (3) The spatial correlations do not obey a power law. That proves that spatially extended systems based on activation-deactivation processes do not have to be point-driven to produce $1/f^\alpha$ noise. The essential ingredient is a local memory of the activation-deactivation process. 
\end{abstract}
]

\narrowtext

A time signal $X(t)$ with zero mean is called \ra or $1/f^\alpha$ signal if its power spectrum $S(f)$ is proportional to $1/f^\alpha$ at low frequencies $f$ with $\alpha \approx 1$. Here, the power spectrum is defined as the amplitude squared of the Fourier transformed signal, i.e.,
\begin{eqnarray}
S(f) = \lim_{T\to\infty} \frac{1}{2T} \left | \int\limits_{-T}^{T} dt X(t) \exp^{-i2\pi f t} \right |^2. \label{powe}
\end{eqnarray}
According to the Wiener-Khinchin theorem, $S(f)$ is the Fourier transform of the autocorrelation function $C(\tau)$ which is defined as
\begin{eqnarray}	
{C} (\tau) = \lim_{T\to\infty} \frac{1}{2T} \int\limits_{-T}^{T} dt X(t + \tau) X(t).
\end{eqnarray}
Consequently, it follows for the autocorrelation function of a signal with $S(f) \propto 1/f^\alpha$ and $0 < \alpha < 1$ that ${C} (\tau) \propto  \left | \tau \right |^{\alpha -1}$. Hence, a $1/f^\alpha$ signal with $\alpha$ close to but smaller than 1 is related to (statistical) long-time correlations which is the reason why \ra is considered to be a particularly interesting phenomenon \emph{a priori}. \\
Moreover, the omnipresence of \ra in nature is one of the oldest puzzles in contemporary physics. It appears in a variety of systems from physics, geophysics, astrophysics, technology, sociology, and biology. Specific examples are the flow of the river Nile \cite{man}, sunspot activity \cite{pres}, pressure variations in the air caused by music and speech \cite{vos}, human coordination \cite{che}, and neuronal spike trains \cite{gru}. One of the most famous examples is the measurement of the voltage drop $V$ on a resistor of resistance $R$ through which a current $I$ is flowing. The power spectrum of the fluctuations around the expected value $V=RI$ clearly shows a $1/f$ behavior over many decades in the frequency domain \cite{wei}.\\
\indent It is natural to expect that there might be a general principle which explains the occurence of $1/f^\alpha$ signals in many of these different systems. However, no generally accepted explanation of the ubiquity of \ra has been proposed yet. Indeed, it is possible to find in the literature some \emph{ad hoc} formulas and theories, but most of them are based on unverified assumptions, or they catch a glimpse of the physics only of some particular system, therefore missing to explain the widespread occurence of the phenomenon (see Refs. \cite{wei,dut} and references therein).\\
\indent In 1987 Bak, Tang, and Wiesenfeld (BTW) introduced the notion of self-organized criticality (SOC) to explain the universality of \ra \cite{btw}. \so systems are nonequilibrium systems driven by their own dynamics to a --- in a statistical sense --- stable state (self-organization). Fluctuations around this state, so-called avalanches, are characterized by power-law distributions in time and space (criticality) implying long-range correlations (for a recent review on SOC see Refs. \cite{bak,jen2}). This automatically leads to a power spectral density exhibiting a $1/f^\alpha$ decay. However, this approach has several shortcomings: First and most important, there is no evidence for power-law space correlations in most systems exhibiting \ra \cite{bla}. This already means that the notion of self-organized criticality and \ra is mutually exclusive in most cases. Second, $\alpha$ is seldom close to one in \so systems. Moreover, the exponent strongly depends on the dimension of the \so system, at least below the upper critical dimension. Finally, as we will show later on, the $1/f^\alpha$ behavior in \so models is observed for high frequencies rather than in the low-frequency range.\\
\indent Recently, several authors have sucessfully modified originally self-organized critical models to overcome these problems \cite{jen,rio,mas,zha}. Despite the diversity of introduced modifications (continuous driving \cite{jen}, dissipation \cite{rio}, (quasi-) one-dimensional geometry \cite{mas,zha}), there is one common denominator. All these models have a preferred propagation direction of the avalanches. This is implicitly defined via specific driving mechanisms. The systems in Refs. \cite{rio,mas,zha} are essentially point driven and the system in Ref. \cite{jen} is boundary driven. Without these special driving mechanisms the models are not able to generate $1/f^\alpha$ noise.\\ 
\indent In this paper, we will show that such a preferred propagation direction is not a necessary condition to obtain \ra from \so models. We propose a simple model with uniform driving able to reproduce the above mentioned characteristics of $1/f^\alpha$ noise.\\

\begin{figure}[bt]
%\centerline{\psfig{figure=graphs/pre1.ps,width=\columnwidth,clip=}}
\centerline{\psfig{figure=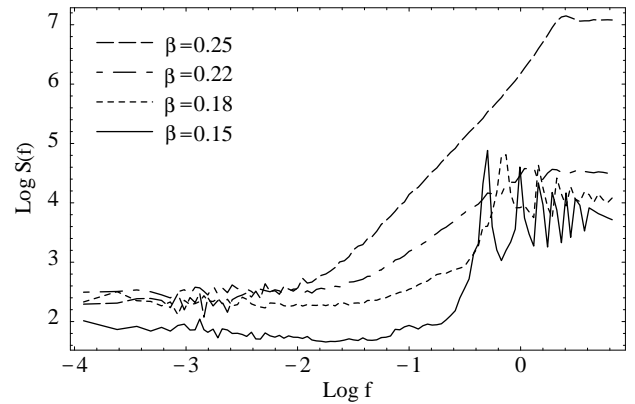,width=\columnwidth,clip=}}
\caption{$\log_{10}$-$\log_{10}$ plot of the power spectrum of $X(t)$ in the two-dimensional OFC model with open boundaries for different $\beta$'s and $N = 2500, F_c = 1, v = 0.1$. There is clearly no sign of $1/f^\alpha$ noise.}
\label{pic1}
\end{figure}

One of the main features believed to be relevant for the description of \ra is an activation-deactivation process \cite{mil}. This is, for example, realized in stick-slip models and, hence, in a model introduced by Olami, Feder and Christensen (OFC) in 1992 which was intended to mimic the dynamics of earthquakes \cite{ofc}. In this model, a real variable $F_i$, called stress, is attached to each point $i$ of a $d$-dimensional cubic lattice of size $N = L^d$. In the initial state, the values of $F$ are randomly distributed in [0,1] obeying a uniform distribution. The dynamic evolution is characterized by slow driving and fast relaxation. All sites $i = 1, \ldots, N$ are driven at the same rate $v$ as long as $F_i < F_c$, i.e.
\begin{eqnarray}
F_i ' = v. \label{driv}
\end{eqnarray}
As soon as one of the $F_i$'s exceeds the critical threshold value $F_c$ the stress $F_i$ is redistributed to the $2d$ nearest neighbors of site $i$,
\begin{eqnarray}
F_i = 0,
\end{eqnarray}
\begin{eqnarray}
F_{nn} = F_{nn} + \beta F_i.
\end{eqnarray}
Here, $\beta$ describes the level of dissipation. The model is conservative for $\beta = 1/2d$ and dissipative for $0 \leq \beta < 1/2d$. The local relaxation continues until all $F_i$'s are subcritical again. The sequence of discharges triggered in this way is called an avalanche. If more than one site is supercritical at any time, the discharges are assumed to happen simultaneously. After the avalanche is over the slow driving [Eq. (\ref{driv})] sets in again. It is important to note that this time scale separation formally implies $v \rightarrow 0$.\\
\indent In two dimensions, the OFC model is considered to be a \so model provided that open boundary conditions are applied. However, it is not clear whether this is only true in the conservative case. Recent investigations seem to imply that in two dimensions the OFC model could only be classified as ``almost critical'' for values of $\beta$ close to but smaller than $0.25$ \cite{car}. This is in contradiction to claims by other groups that the model is self-organized critical even in a certain range of dissipative values of $\beta$ \cite{ofc,gra,mid}. However, the distribution of avalanches with respect to their size obeys a power law in the numerically accessable range of system sizes even for a small amount of dissipation.\\
\indent The model, as it stands, is not a good candidate to describe \ra as follows from Fig. \ref{pic1}. There the power spectrum of the OFC model is shown for different values of $\beta$. The quantity we use as our time signal is the avalanche signal
\begin{eqnarray}
X(t) = \sum_{j} g_j \delta(t-t_j),
\end{eqnarray}
where $g_j$ denotes the size (i.e., the number of topplings) of the $j$th avalanche and $t_j$ its time of occurence on the time scale of the driving. The explicit definition of a time scale leads to a time signal with varying temporal distances between avalanches. This is in contrast to other signals considered so far in the context of \so and $1/f^\alpha$ noise. The $\delta$ function is motivated by the time scale separation, i.e., we are not able to observe events on the time scale of the relaxation. This is also reasonable because we are only interested in the low-frequency range where \ra is usually found.\\
\indent The time signal was recorded after the system had reached a stationary state as described in Ref. \cite{gra}. For nonzero dissipation, a characteristic frequency occurs in the spectrum as already discussed in Refs. \cite{gra,jan}. Above and especially below the characteristic frequency, there is clearly no sign of $1/f^\alpha$ noise. Rather a white noise behavior can be identified.\\
This is not in contradiction to the observations in Refs. \cite{jan,chr}. In Ref. \cite{jan}, a $1/f^2$ behavior was found for frequencies larger than the characteristic frequency. However, a different time signal was used, namely a stress signal which is the stress averaged over the lattice sites as a function of time. In Ref. \cite{chr}, a $1/f$-type behavior was described. The authors measured the avalanche signal in terms of the time scale of the relaxation, i.e., the time between different avalanches was essentially set to zero. Hence, they observed a high-frequency phenomenon characterizing the internal temporal development of the avalanches. Therefore, their findings cannot be considered as $1/f^\alpha$ noise. As we will see later, our modifications lead to a $1/f$ decay in the power spectrum below the characteristic frequency which is the range where \ra should be looked for.\\
\indent Our model is basically an extension of the OFC model. We just add a single new element: The threshold value $F_c$ becomes a function of space and time mimicking a local memory such that each site remembers its (cummulative) history of discharges. The simplest way to model such a memory is to implement it by a random process. After each toppling the respective $F_c(i)$ evolves according to a random walk with Gaussian step length
\begin{eqnarray}
F_{c, \tau} (i) = F_{c, \tau-1} (i) + \sigma \eta(i,\tau),
\end{eqnarray}
where $\tau$ denotes the number of topplings of site $i$ and $\{\eta(i,\tau)\}$ the sequence of uncorrelated normally distributed random variables with zero expectation and unit variance. The strength of the white noise source is given by $\sigma$. To omit negative threshold values and to confine the random walk, we impose reflecting boundaries at $0$ and at $F_u = 2 F_c$. As initial condition, we use $F_{c, 0} (i) = F_c$.\\

\begin{figure}[bt]
%\centerline{\psfig{figure=graphs/pre2.ps,width=\columnwidth,clip=}}
\centerline{\psfig{figure=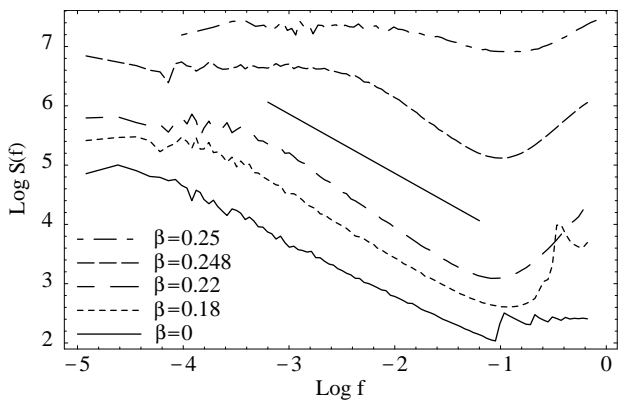,width=\columnwidth,clip=}}
\caption{$\log_{10}$-$\log_{10}$ plot of the power spectrum of $X(t)$ in our model with open boundaries for different $\beta$'s and $\sigma = 0.04, d = 2, N = 2500, F_c = 1, v = 0.1$. The solid line with exponent $1.0$ is drawn for reference. The dissipative version of the model clearly generates $1/f^\alpha$ noise. For $d = 1$ as well as for larger system sizes, we obtain similar results.}
\label{mod1}
\end{figure}

\begin{figure}[bt]
%\centerline{\psfig{figure=graphs/pre3.ps,width=\columnwidth,clip=}}
\centerline{\psfig{figure=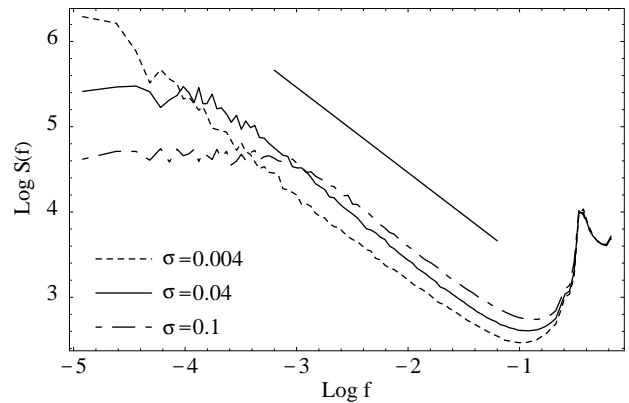,width=\columnwidth,clip=}}
\caption{$\log_{10}$-$\log_{10}$ plot of the power spectrum of $X(t)$ in our model with open boundaries for different $\sigma$'s and $\beta = 0.18, d = 2, N = 2500, F_c = 1, v = 0.1$. The solid line with exponent $1.0$ is drawn for reference. For $d = 1$ as well as for larger system sizes, we obtain similar results.}
\label{mod2}
\end{figure}

\begin{figure}[bt]
%\centerline{\psfig{figure=graphs/pre4.ps,width=\columnwidth,clip=}}
\centerline{\psfig{figure=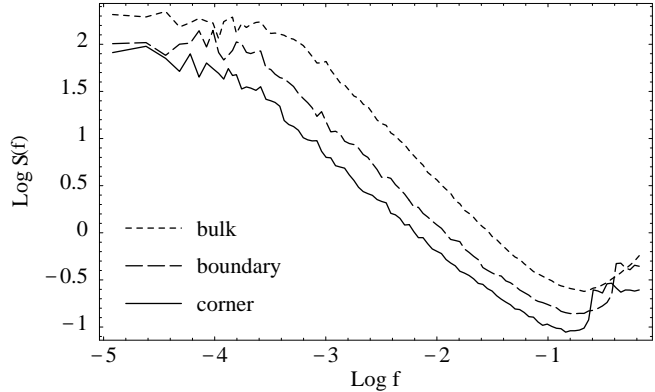,width=\columnwidth,clip=}}
\caption{$\log_{10}$-$\log_{10}$ plot of the power spectrum of $X(t,i)$ in our model with open boundaries for different sites $i$ and $\beta = 0.22, \sigma = 0.04, d = 2, N = 2500, F_c = 1, v = 0.1$. For bulk sites, $\alpha = 1.2$ which is exactly the same as for $X(t)$. For boundary sites, $\alpha = 1.1$.}
\label{pic3}
\end{figure}

Computing the power spectrum for our model, we find a clear $1/f^\alpha$ decay over several decades for dissipative $\beta$'s both in one and two dimensions (see Figs. \ref{mod1} and \ref{mod2}). The exponent decreases slightly with decreasing $\beta$ and lies between 0.9 and 1.2. As a rule, the range of the $1/f^\alpha$ behavior also decreases with decreasing dissipation shrinking to zero in the conservative limit. In this limit, the crossover leads to a white-noise type of behavior at low frequencies.
Our modification also destroys the power-law distribution of the avalanches in the dissipative case. We find an exponential distribution instead. Hence, avalanches cannot establish long-range correlations anymore through the system. This is expected since the responsible mechanism for the critical or ``almost critical'' behavior is marginal synchronization \cite{gra,mid}. This synchronization is necessarily destroyed as soon as the threshold varies locally (for quenched random thresholds see Ref. \cite{jan}).\\
\indent Extensive numerical simulations show that our results are very stable with respect to variations in the parameters. Different values of $\sigma$ and $F_u$ lead to the same results as long as $F_u \gg \sigma \gg 0$ (see Fig. \ref{mod2}). This is true for different types of distributions of the random increments $\{\eta(i,\tau)\}$, too. Except for the transients, the behavior of the model is also independent of the initial distribution of the $F_{c, 0} (i)$'s on the interval $[0, F_u]$. Our findings do not depend on the choice of boundary conditions as well. Periodic and open boundaries give similar power spectra. Even variations in the dynamic rules of the present model as realized, for example, in the Feder-Feder model \cite{fed} do not alter our results. This points towards a generic behavior strongly supporting the view in Refs. \cite{rio,mil} that nonlinearity (here, activation/deactivation of sites with evolving thresholds) and dissipation are among the relevant features for generating $1/f^\alpha$ noise.\\
\indent The explanation of our results is the following: Due to the absence of critical behavior, only small avalanches occur in our model for dissipative $\beta$'s. Consequently, one can think of $S(f)$ in a first approximation as the superposition of local power spectra $S(f,i)$,
\begin{eqnarray}
S(f) \approx \sum\limits_{i=0}^{N} S(f,i). \label{supe}
\end{eqnarray}
The local signal $X(t,i)$ generating the respective $S(f,i)$ is just the avalanche signal at site $i$, i.e., the number of topplings of this site during an avalanche at time $t$. This means that the sum over $i$ of the $X(t,i)$ is just $X(t)$.\\
\indent We have investigated the local power spectra and we find indeed that Eq. (\ref{supe}) is a good approximation. This underlines especially that there is no dependence on the dimension of the system. It turns out that the $S(f,i)$ are almost independent of $i$ and that they show a $1/f^\alpha$ behavior themselves (see Fig. \ref{pic3}).\\

This result can be understood to a certain extent by mapping our model to a model introduced by Kaulakys and Me\v{s}kauskas \cite{kau}. In order to do so, we have to neglect all interactions between different sites. This means considering the limit $\beta = 0$ or, equivalently, $N=1$. Additionally, the random walk of the threshold is no longer confined by two reflecting boundaries. A parabolic potential centered around $F_c$ and characterized by the relaxation rate $\gamma$ is used instead, 
\begin{eqnarray}
F_{c, \tau} =  F_c + \Delta F_{c, \tau},
\end{eqnarray}
\begin{eqnarray}
\Delta F_{c, \tau} = (1 - \gamma) \Delta F_{c, \tau-1} + \sigma \eta(\tau),
\end{eqnarray}
with $\Delta F_{c, 0} = 0$. Since we consider $N=1$, the time signal $X(t)$ simplifies considerably: $g_j = 1$ for all $j$ and, due to the uniform driving, the $t_j$'s are given by
\begin{eqnarray}
t_j = t_{j-1} + \frac{F_{c,j}}{v},
\end{eqnarray}
\begin{eqnarray}
F_{c,j} = F_{c,j-1} - \gamma (F_{c,j-1} - F_c) + \sigma \eta(j),
\end{eqnarray}
with $F_{c,1} = \frac{F_{c}}{v}$ and $t_0 = 0$. Hence, $X(t)$ is already determined by the series of $\Delta t_j = t_j - t_{j-1}$ which evolve according to a random walk in a parabolic potential. This corresponds to one particle moving in a closed contour with the period of the drift of the particle around the contour fluctuating about the average value $\frac{F_c}{v}$. The $t_j$'s are then the transit times measured at a certain point. 
\indent This is indeed the model introduced in Ref. \cite{kau}. Kaulakys and Me\v{s}kauskas (KM) were able to compute analytically the power spectral density and obtained a power law with $\alpha=1$. This behavior can be found in any desirably wide range of frequencies for a sufficiently small $\gamma$. The crucial point is that the law of large numbers is not valid for $\lim_{n \to \infty} \frac{1}{n} \sum_{j=0}^{n-1} \Delta t_j $ (see the Appendix).\\
Their results explain at least qualitatively our findings in the limit $\beta = 0$. However, the exponent $\alpha$ differs. In our model, we find $\alpha = 0.91 \pm 0.02$ for $\beta = 0$. This is due to an important difference between the KM model and our model. As already noted, $\gamma$ has to be small to generate $1/f^\alpha$ noise. Moreover, the asymptotic distribution of the $\Delta t$'s is a Gaussian with mean $0$ and variance $\frac{\sigma^2}{2 \gamma}$. This means that there is a non neglectable probability of negative $\Delta t$'s. Consequently, $t_{j+1}$ can be smaller than $t_j$ implying a causality backwards in time. Hence, the KM model is somewhat ill-defined. In our model, negative $\Delta t$'s are not possible since the threshold has to be larger than or equal to zero. This also implies that the $1/f$ behavior cannot be extended to any desirable wide range of frequencies.\\

In conclusion, we have shown that uniform driven SOC models are generally not able to generate \ra without further modifications. The essential ingredient that has to be added is a local memory. This proves that spatially extended systems based on activation-deactivation processes do not have to be point driven to produce $1/f^\alpha$ noise.\\
\indent In the present model, the local memory is realized in the easiest possible way by a random walk of the threshold. We showed that the dynamics of the threshold is equivalent to the KM model of transit times under certain assumptions in the limit $\beta \rightarrow 0$. This means that our model can be considered as a physically reasonable generalization of the KM model to systems with a threshold - even if they are spatially extended. As a consequence, the present model combines the idea that \ra may result from the clustering of the signal pulses \cite{kau} with the view that an activation-deactivation process and dissipation are the main features relevant for the description of \ra \cite{rio,mil}. The robustness of our results strongly supports these views.\\
\indent As an experimental realization of our model, we suggest a stick-slip system with a Markovian threshold evolution. Finally, we would like to point out that the present model is similar to the coupled ``integrate-and-fire'' oscillators studied in the context of neuronal networks and biology. Work is in progress to investigate these connections further and might lead to an explanation for the occurence of \ra in cortical neurons \cite{gru}. 

%\acknowledgments

J. Davidsen would like to thank the Land Schleswig-Holstein, Germany, for financial support.

\appendix

\section*{Computation of the power spectral density}
Consider a signal $X(t) = \sum_{j} \delta(t-t_j)$ as in the KM model. Define $\Delta t_j = t_j - t_{j-1}$. It follows for the power spectral density [see Eq. (\ref{powe})]:
\begin{eqnarray}
S(f) &=& \lim_{T \rightarrow \infty} \frac{1}{2T} \left| \sum_j \exp^{-i2\pi f t_j}\right|^2, \\
&=& \lim_{T \rightarrow \infty} \frac{1}{2T}  \sum_j \sum_q \exp^{i2\pi f (t_{j+q}-t_j)}, \\
&=& \lim_{T \rightarrow \infty} \frac{1}{2T}  \sum_j \sum_q \exp^{i2\pi f q \sum_k \frac{\Delta t_k}{q}}.
\end{eqnarray}
With $\bar{I} = \lim_{T \rightarrow \infty} \frac{1}{2T} (j_{max} - j_{min} + 1)$, this leads to
\begin{eqnarray}
S(f) &=& \bar{I} \left\langle \sum_q \exp^{i2\pi f q \sum_k \frac{\Delta t_k}{q}} \right\rangle,
\label{equkm}
\end{eqnarray}
where $\langle ... \rangle$ denotes the average over the ensemble and over j. Hence, all we need is the probability distribution $\Psi$ of the $1/q \sum_{k=0}^{q-1} \Delta t_k$. For the KM model with an average period $\Delta t$, it was shown that $\Psi$ is a Gaussian with mean $\Delta t$ and variance $\frac{\sigma^2}{2 \gamma}$ for $f > \gamma^{3/2} / \pi \sigma$ \cite{kau}. Hence, $1/q \sum_k \Delta t_k$ obeys the same distribution as $\Delta t_j$ and does not depend on $q$. This can be used to further simplify Eq. (\ref{equkm}):
\begin{eqnarray}
S(f) &=& \bar{I} \sum_q \left\langle \exp^{i2\pi f q \sum_k \frac{\Delta t_k}{q}} \right\rangle. 
\end{eqnarray}
For small enough $f$, the summation can be replaced by an integral. (In the KM model, this is valid for $f <  2 \sqrt{\gamma} / \pi \sigma$ and $f \ll (2 \pi \Delta t)^{-1}$.) Changing variables from $q$ to $q' = qf$ and evaluating the integrals gives
\begin{eqnarray}
S(f) = \bar{I} \Psi(0)/f.
\end{eqnarray}
The fact that the probability distribution of $\frac{1}{q} \sum_{k=0}^{q-1} \Delta t_k$ does not depend on $q$ merely means that the law of large numbers is not valid due to the correlation between different $\Delta t_k$'s.\\
This is of course different in the case of independent random variables $\Delta t_k$, i.e. for $\gamma = 1$. In this case the distribution of $\frac{1}{q} \sum_{k=0}^{q-1} \Delta t_k$ does depend on $q$. To be more precise, the distribution is still a Gaussian with the same mean as before but the variance becomes $\frac{\sigma^2}{2 q}$. In the limit $q \rightarrow \infty$, the variance vanishes. A short calculation gives as expected
\begin{eqnarray}
S (f) \propto \frac{1}{f^2 + f_0}.
\end{eqnarray}
This is exactly what happens in the KM model for $f > \gamma^{3/2} / \pi \sigma$ \cite{kau}.\\
\indent To summarize, the crucial ingredient to obtain $1/f$ noise is the generation of strong enough correlations between different $\Delta t_j$ such that $1/q \sum_k \Delta t_k$ obeys the same distribution as $\Delta t_k$ does. The specific form of $\Psi$ is not important as long as $\Psi (0) \neq 0$. In the KM model, the strong correlations are implemented by a random walk dynamics. In general, other stochastic mechanisms are capable of generating such correlations as well, e.g., shot noise in combination with fast relaxation giving rise to random flows of events without memory and Cauchy statistics \cite{kuz1,kuz2}.

\end{document}